\documentclass[
  ,final            
  ]
  {aipproc}

\layoutstyle{6x9}

\usepackage{epsfig}
\usepackage{wrapfig,rotating}
\usepackage{amssymb}
\usepackage{amsmath}
\usepackage{hyperref}

\newcommand{\mbf}[1]{\mathbf{#1}}

\begin{document}

\title{AdS/QCD and its Holographic Light-Front  \\
Partonic Representation 
}


  \author{Guy F. de T\'eramond}{
  address={Universidad de Costa Rica, San Jos\'e, Costa Rica}
  }

\author{Stanley J. Brodsky}{
  address={SLAC National Accelerator Laboratory, 
Stanford University, Stanford, CA 94309, USA }
  }

\begin{abstract}

Starting from the Hamiltonian equation of motion in QCD
we find a single variable light-front equation for QCD which determines the eigenspectrum
and the light-front wavefunctions of hadrons for general spin and
orbital angular momentum. This light-front wave equation is
equivalent to the
equations of motion which describe the propagation of spin-$J$ modes
on  anti-de Sitter (AdS) space. 

\end{abstract}

\maketitle

\section{Introduction}

Quantum Chromodynamics, the Yang-Mills local gauge field theory of
$SU(3)_C$ color symmetry, provides a fundamental description of
hadronic physics in terms of quark and gluon degrees of
freedom.  Yet,  because of the 
strongly coupling nature of QCD in the infrared domain,
it has been difficult to determine how the constituents
appear in the physical spectrum as colorless states, mesons and baryons, 
or to make precise predictions for hadronic properties outside of the perturbative regime.  Thus an important theoretical objective is to find an initial approximation to relativistic bound state problems in QCD which is analytically tractable and which can be systematically improved.

Light-front quantization is the ideal framework to describe the
structure of hadrons in terms of their quark and gluon degrees of
freedom.  The simple vacuum structure in the light-front (LF) 
allows unambiguous definition of the partonic content of a hadron: partons 
in a hadronic state are described by light-front wave functions (LFWFs)
which encode the hadronic properties. 
The LFWFs of bound states in QCD are
relativistic generalizations of the Schr\"odinger wavefunctions of
atomic physics, but they are determined at fixed light-cone time
$\tau  = t +z/c$ -- the ``front form'' introduced by
Dirac~\cite{Dirac:1949cp} -- rather than at fixed ordinary time $t$.

In this talk, we show that there is an invariant
light-front coordinate $\zeta$ which allows the separation of the
essential dynamics of quark and gluon binding from the kinematical
physics of the constituents. The result is a single-variable LF equation for QCD
which determines the spectrum and the LFWFs
of hadrons for general spin and orbital angular momentum.
Our analysis follows from recent developments in light-front
QCD~\cite{deTeramond:2008ht,Brodsky:2006uqa,Brodsky:2007hb,Brodsky:2008pg,Brodsky:2008td}
which have been inspired by the
AdS/CFT correspondence~\cite{Maldacena:1997re} between string states
in anti-de Sitter (AdS) space and conformal field theories (CFT) in
physical space-time. The use of AdS space and conformal
methods in QCD can be motivated from the empirical
evidence~\cite{Deur:2008rf} and theoretical
arguments~\cite{Brodsky:2008be} that the QCD coupling $\alpha_s(Q^2)
$ has an infrared fixed point at low $Q^2.$  As
we have shown recently, there is a remarkable mapping between the
description of hadronic modes in AdS space and the Hamiltonian
formulation of QCD in physical space-time quantized on the
light-front~\cite{deTeramond:2008ht}. This procedure allows string modes $\Phi(z)$ in the AdS
holographic variable $z$ to be precisely mapped to the LFWFs  of hadrons 
in physical space-time in terms of a
specific LF variable $\zeta$ which measures the separation
of the quark and gluonic constituents within the hadron. This mapping was
originally obtained by matching the expression for electromagnetic
current matrix elements in AdS space with the corresponding
expression for the current matrix element using LF theory
in physical space time~\cite{Brodsky:2006uqa,Brodsky:2007hb}. More recently we have
shown that one obtains a consistent holographic mapping using the
matrix elements of the energy-momentum tensor~\cite{Brodsky:2008pf},
thus providing an important verification of
holographic mapping from AdS to physical observables defined on the
light front.

\section{A Single-Variable Light-Front Equation for QCD}

To  a first approximation
light-front QCD is formally equivalent to an effective gravity
theory on AdS$_5$. To prove this, we show that  the LF Hamiltonian
equation of motion of QCD leads to an effective LF wave equation for
physical modes  $\phi(\zeta)$ which encode the hadronic properties.
We compute the hadron mass $\mathcal{M}^2$ from the hadronic matrix element
$\langle \psi_H(P') \vert H_{LF}\vert\psi_H(P) \rangle  \! = \!
\mathcal{M}_H^2  \langle \psi_H(P' ) \vert\psi_H(P) \rangle$, 
where $H_{LF}$ is the Lorentz invariant Hamiltonian
$H_{LF}= P_\mu P^\mu = P^-P^+  \! - \mbf{P}^2_\perp$ and
the state $\vert \psi_H \rangle$ is an expansion 
in multi-particle Fock states
$\vert n \rangle $ of the free LF Hamiltonian:
~$\vert \psi_H \rangle = \sum_n \psi_{n/H} \vert n \rangle$.
To simplify the discussion we will consider a two-parton hadronic
bound state in the limit of massless constituents. We find ~\cite{deTeramond:2008ht}
\begin{eqnarray} \nonumber
\mathcal{M}^2  &\!\!=\!\!&  \int_0^1 \! d x \! \int \!  \frac{d^2
\mbf{k}_\perp}{16 \pi^3}   \,
  \frac{\mbf{k}_\perp^2}{x(1-x)}
 \left\vert \psi (x, \mbf{k}_\perp) \right \vert^2  + {\rm interactions} \\  \label{eq:Mb}
  &\!\!=\!\!& \int_0^1 \! \frac{d x}{x(1-x)} \int  \! d^2 \mbf{b}_\perp  \,
  \psi^*(x, \mbf{b}_\perp)
  \left( - \mbf{\nabla}_{ {\mbf{b}}_{\perp \ell}}^2\right)
  \psi(x, \mbf{b}_\perp)   +  {\rm interactions}.
 \end{eqnarray}
The functional dependence  for a given Fock state is
given in terms of the invariant mass
$
 \mathcal{M}_n^2  = \left( \sum_{a=1}^n k_a^\mu\right)^2 
 \to \frac{\mbf{k}_\perp^2}{x(1-x)} \,,
$
the measure of the off-mass shell energy~ $\mathcal{M}^2 \! - \mathcal{M}_n^2$.
Similarly in impact space the relevant variable is  $\zeta^2= x(1\!-\!x)\mbf{b}_\perp^2$  for a two-parton state.
Thus, to first approximation  LF dynamics  depend only on the boost
invariant variable $\mathcal{M}_n$ or $\zeta$ and hadronic
properties are encoded in the hadronic mode $\phi(\zeta)$:
 $ \psi(x, \mbf{k}_\perp) \to \phi(\zeta)$.
 We choose the normalization of  the LF mode $\phi(\zeta) = \langle \zeta \vert \phi \rangle$ with
$
 \int \! d \zeta \, \vert \langle \zeta \vert \phi\rangle\vert^2 = 1.
$
 Comparing with the LFWF normalization, we find the functional relation:  
 $\frac{\vert \phi \vert^2}{\zeta} = \frac{2 \pi}{x(1-x)} \vert\psi(x, \mbf{b}_\perp)\vert^2$. 
 
We  write the Laplacian operator in circular
cylindrical coordinates $(\zeta, \varphi)$, and factor
out the angular dependence of the modes in terms of the $SO(2)$
Casimir representation $L^2$ of orbital angular momentum in the
transverse plane: $\phi(\zeta, \varphi) \sim e^{\pm i L
\varphi} \phi(\zeta)$. We find
\begin{equation}
\mathcal{M}^2  = \int \! d\zeta \, \phi^*(\zeta) \left(
-\frac{d^2}{d\zeta^2} - \frac{1 - 4L^2}{4\zeta^2} \right)
\phi(\zeta) +  \int \! d\zeta \, \phi^*(\zeta) U(\zeta) \phi(\zeta) ,
\end{equation}
where all the complexity of the interaction terms in the QCD
Lagrangian is summed up in the effective potential $U(\zeta)$. The
light-front eigenvalue equation $H_{LF} \vert \phi \rangle =
\mathcal{M}^2 \vert \phi \rangle$ is thus a light-front wave
equation for $\phi$
\begin{equation} \label{eq:QCDLFWE}
\left(-\frac{d^2}{d\zeta^2} - \frac{1 - 4L^2}{4\zeta^2} + U(\zeta)
\right) \phi(\zeta) = \mathcal{M}^2 \phi(\zeta),
\end{equation}
an effective single-variable light-front Schr\"odinger equation
which is relativistic, covariant and analytically tractable. One can readily generalize the equations to allow for
the kinetic energy of massive quarks~\cite{Brodsky:2008pg}.

As the simplest example we consider a bag-like
model where the partons are free inside the
hadron and the interaction terms will effectively build confinement.
The effective potential is a hard wall: $U(\zeta) = 0$ if  $\zeta
\le \frac{1}{\Lambda_{\rm QCD}}$ and
 $U(\zeta) = \infty$ if $\zeta > \frac{1}{\Lambda_{\rm QCD}}$.
 If $L^2 \ge 0$ the LF Hamiltonian is positive definite
 $\langle \phi \vert H_{LF} \vert \phi \rangle \ge 0$ and thus $\mathcal M^2 \ge 0$.
 If $L^2 < 0$ the LF Hamiltonian is unbounded from below and the particle
 ``falls towards the center''. The critical value corresponds to $L=0$.
  The mode spectrum  follows from the boundary conditions
 $\phi \! \left(\zeta = 1/\Lambda_{\rm QCD}\right) = 0$, and is given in
 terms of the roots of Bessel functions: $\mathcal{M}_{L,k}^2 = \beta_{L, k} \Lambda_{\rm QCD}$.
 Since in the conformal limit $U(\zeta) \to 0$, the hard-wall LF model discussed here is equivalent to the AdS/CFT
 hard wall model of
 Ref.~\cite{Polchinski:2001tt}. Likewise a two-dimensional transverse oscillator with
 effective potential $U(\zeta) \sim \zeta^2$ is equivalent to the soft-wall model of
 Ref.~\cite{Karch:2006pv} which reproduce the usual linear Regge trajectories.
 Upon the substitution $\zeta \! \to\! z$  and
$\Phi_J(z)    \!  \sim \! (z/R)^{3/2 - J} \phi(z)$
in (\ref{eq:QCDLFWE}) we find the equation of motion
\begin{equation} \label{eq:eomPhiJz}
\left[ z^2 \partial_z^2 - (d\! -\! 1 \!- \!2 J) z \, \partial_z + z^2 \mathcal{M}^2
\!  -  (\mu R)^2 \right] \!  \Phi_J  = 0,
\end{equation}
describing the propagation of a spin-$J$ mode in AdS$_{d+1} $space.
For $d=4$ the fifth dimensional mass $ (\mu R)^2 = - (2-J)^2 + L^2$.
The scaling dimensions are $\Delta = 2 + L$ independent of $J$ in agreement with the
twist scaling dimension of a two parton bound state in QCD.

\section{Transition Matrix Elements}

 Light-Front Holography can be derived by observing the correspondence between matrix elements obtained in AdS/CFT with the corresponding formula using the LF 
representation~\cite{Brodsky:2006uqa} .  The light-front electromagnetic form factor in impact 
space~\cite{Brodsky:2006uqa,Brodsky:2007hb} can be written as a sum of overlap of light-front wave functions of the $j = 1,2, \cdots, n-1$ spectator
constituents:
\begin{equation} \label{eq:FFb} 
F(q^2) =  \sum_n  \prod_{j=1}^{n-1}\int d x_j d^2 \mbf{b}_{\perp j}  \sum_q e_q
\exp \! {\Bigl(i \mbf{q}_\perp \! \cdot \sum_{j=1}^{n-1} x_j \mbf{b}_{\perp j}\Bigr)} 
\left\vert \psi_n(x_j, \mbf{b}_{\perp j})\right\vert^2.
\end{equation}
The formula is exact if the sum is over all Fock states $n$.
For definiteness we shall consider a two-quark $\pi^+$  valence Fock state 
$\vert u \bar d\rangle$ with charges $e_u = \frac{2}{3}$ and $e_{\bar d} = \frac{1}{3}$.
For $n=2$, there are two terms which contribute to the $q$-sum in (\ref{eq:FFb}). 
Exchanging $x \leftrightarrow 1-x$ in the second integral  we find ($e_u + e_{\bar d}$ = 1)
\begin{equation}  \label{eq:PiFFb}
 F_{\pi^+}(q^2) 
=  2 \pi \int_0^1 \! \frac{dx}{x(1-x)}  \int \zeta d \zeta\, 
J_0 \! \left(\! \zeta q \sqrt{\frac{1-x}{x}}\right) 
\left\vert \psi_{u \bar d/ \pi}\!(x,\zeta)\right\vert^2,
\end{equation}
where $\zeta^2 =  x(1-x) \mathbf{b}_\perp^2$ and $F_\pi^+(q\!=\!0)=1$. 
We now compare this result with the electromagnetic form-factor in AdS space:
$
F(Q^2) = R^3 \int \frac{dz}{z^3} \, J(Q^2, z) \vert \Phi(z) \vert^2,
$
where $J(Q^2, z) = z Q K_1(z Q)$.
Using the integral representation 
$J(Q^2, z) = \int_0^1 \! dx \, J_0\negthinspace \left(\negthinspace\zeta Q
\sqrt{\frac{1-x}{x}}\right)$,
we can write the AdS electromagnetic form-factor as
\begin{equation} 
F(Q^2)  =    R^3 \! \int_0^1 \! dx  \! \int \frac{dz}{z^3} \, 
J_0\!\left(\!z Q\sqrt{\frac{1-x}{x}}\right) \left \vert\Phi(z) \right\vert^2 .
\label{eq:AdSFx}
\end{equation}
Comparing with the light-front QCD  form factor (\ref{eq:PiFFb}) for arbitrary  values of $Q$
 we find a functional relation between the light-front wave function $\psi(x,\zeta)$ and the AdS wavefunction $\Phi(z)$
consistent with the results found in the previous section from the QCD light-front Hamiltonian eigenvalue equation.
We identify the transverse light-front variable $\zeta$, $0 \leq \zeta \leq \Lambda_{\rm QCD}$,
with the holographic variable $z$.

\section{Conclusion}

\begin{figure}[htbp] 
\includegraphics[width=11.5cm]{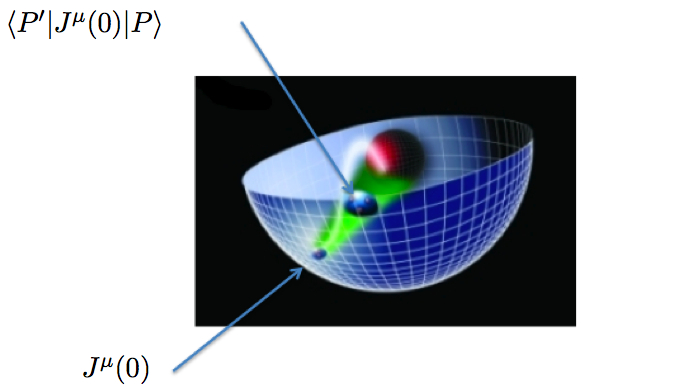}
\caption{AdS representation of light-front QCD.   Different values of $z$ correspond to different scales at which
the hadron is examined. The AdS boundary at $z \to 0$  (large circumference) corresponds to the $Q \to \infty$, UV, zero separation limit. Local operators are defined at the AdS boundary;
hadronic transition matrix elements probe the wave function $\Phi(z)$ (also represented in the figure) at  $z = 1/Q$. The inner sphere at large $z$ represents the IR confinement radius.  In the figure, a small proton created  by a local interpolating
operator at the AdS boundary falls into AdS space pulled by the gravitational field up to its larger size allowed by confinement. Due to the warp factor the proton size shrinks by a factor $z/R$ as observed in Minkowski space.}
\label{default}
\label{Fig1}
\end{figure}

We have shown that the use of the invariant coordinate $\zeta$ in light-front QCD
allows the separation of the dynamics of quark and gluon binding
from the kinematics of constituent spin and internal
orbital angular momentum. The result is a single-variable LF
Schr\"odinger equation  which determines the spectrum
and  LFWFs of hadrons for general spin and
orbital angular momentum. 
This LF wave equation serves as a first approximation to QCD and is equivalent to the
equations of motion which describe the propagation of spin-$J$ modes
on  AdS. 
This allows us to establish
a gauge/gravity correspondence between an effective gravity theory
on AdS$_5$ and light front QCD.
Remarkably, the AdS equations
correspond to the kinetic energy terms of  the partons inside a
hadron, whereas the interaction terms build confinement and
correspond to the truncation of AdS space. 
Identical results are found by matching the expression of current matrix elements or the 
energy-momentum tensor in AdS space with the corresponding expressions 
using light-front theory in physical space time. This is illustrated in Fig \ref{Fig1}, where
local operators, such as the current  $J^\mu(0)$, are defined at the asymptotic AdS  boundary at $z \to 0$
(large circumference)  whereas hadronic transition matrix elements like $\langle P'\vert J^\mu(0) \vert P \rangle$
probe the hadronic wave function $\Phi(z)$ at a distance $z = 1/Q$, $Q = P' - P$, inside AdS space.

One can systematically improve the holographic approximation by
diagonalizing the QCD light-front Hamiltonian on the AdS/QCD basis.
The action of the non-diagonal terms
in the QCD interaction Hamiltonian generates the form of the higher
Fock state structure of hadronic LFWFs. We emphasize, that in
contrast with the original AdS/CFT correspondence, the large $N_C$
limit is not required to connect light-front QCD to
an effective dual gravity approximation.

\section*{Acknowledgments}

Presented by GdT at Diffraction 2008, International Workshop on Diffraction in High Energy 
Physics, La Londe-les-Maures, France, 9-14 September  2008.  He thanks the organizers of the meeting for their invitation and in particular Jacques Soffer for his outstanding hospitality.  This research was supported by the Department
of Energy contract DE--AC02--76SF00515 and by Fondo de Incentivos CONICIT/MICIT Costa Rica. SLAC-PUB-13457.


\begin{thebibliography}{99}

 \bibitem{Dirac:1949cp}
  P.~A.~M.~Dirac,
  Rev.\ Mod.\ Phys.\  {\bf 21}, 392 (1949).
  
  \bibitem{deTeramond:2008ht}
  G.~F.~de Teramond and S.~J.~Brodsky,
  arXiv:0809.4899 [hep-ph].

\bibitem{Brodsky:2006uqa}
  S.~J.~Brodsky and G.~F.~de Teramond,
  Phys.\ Rev.\ Lett.\  {\bf 96}, 201601 (2006)
  [arXiv:hep-ph/0602252].
  
   \bibitem{Brodsky:2007hb}
  S.~J.~Brodsky and G.~F.~de Teramond,
  Phys.\ Rev.\  D {\bf 77}, 056007 (2008)
  [arXiv:0707.3859 [hep-ph]].

\bibitem{Brodsky:2008pg}
  S.~J.~Brodsky and G.~F.~de Teramond,
  arXiv:0802.0514 [hep-ph].
  
  \bibitem{Brodsky:2008td}
  S.~J.~Brodsky and G.~F.~de Teramond,
  arXiv:0810.1876 [hep-ph].
  
  \bibitem{Maldacena:1997re}
  J.~M.~Maldacena,
  Adv.\ Theor.\ Math.\ Phys.\  {\bf 2}, 231 (1998)
  [Int.\ J.\ Theor.\ Phys.\  {\bf 38}, 1113 (1999)]
  [arXiv:hep-th/9711200].

\bibitem{Deur:2008rf}
  A.~Deur, V.~Burkert, J.~P.~Chen and W.~Korsch,
  Phys.\ Lett.\  B {\bf 665}, 349 (2008)
  [arXiv:0803.4119 [hep-ph]].

\bibitem{Brodsky:2008be}
  S.~J.~Brodsky and R.~Shrock,
  Phys.\ Lett.\  B {\bf 666}, 95 (2008)
  [arXiv:0806.1535 [hep-th]].
  
  \bibitem{Brodsky:2008pf}
  S.~J.~Brodsky and G.~F.~de Teramond,
  Phys.\ Rev.\  D {\bf 78}, 025032 (2008)
  [arXiv:0804.0452 [hep-ph]].
  
  
  \bibitem{Polchinski:2001tt}
  J.~Polchinski and M.~J.~Strassler,
  Phys.\ Rev.\ Lett.\  {\bf 88}, 031601 (2002).

 \bibitem{Karch:2006pv}
  A.~Karch, E.~Katz, D.~T.~Son and M.~A.~Stephanov,
  Phys.\ Rev.\  D {\bf 74}, 015005 (2006).
  
  
\end{thebibliography}
\end{document}